\documentclass[11pt]{article}

\usepackage[left=1.5cm, right=1.5cm, top=3.0cm, bottom=3.0cm]{geometry}
\usepackage{graphicx,color}
\usepackage{array,tabularx,booktabs,geometry,subfigure,graphicx,longtable}
\usepackage{amsmath,amssymb,amsfonts}
\usepackage{extarrows}
\usepackage[hyperfootnotes=false, linktocpage=true, colorlinks, citecolor=blue, linkcolor=blue, urlcolor=blue]{hyperref}
\usepackage[all]{xy} 
\usepackage[mathscr]{eucal}
\usepackage{hypbmsec}
\usepackage{fancyvrb}
\usepackage{flexisym}
\usepackage{float}
\usepackage{fixltx2e,mathtools}
\usepackage{cite}
\usepackage[bf]{caption}
\MakeRobust{\eqref}

\newcommand{\IP}{\mathbb{P}}

\providecommand{\id}{\leavevmode\hbox{\small$\mathrm{1}$\kern-3.8pt\normalsize$\mathrm{1}$}}
\def\fnote#1#2{\begingroup\def\thefootnote{#1}\footnote{#2}
     \addtocounter{footnote}{-1}\endgroup}

\begin{document}

\vspace{1cm}

\title{       {\Large \bf Numerical Metrics, Curvature Expansions\\ and Calabi-Yau Manifolds}}

\vspace{2cm}

\author{
Wei~Cui${}{}$ and
James~Gray${}{}$
}
\date{}
\maketitle
\begin{center} {\small ${}${\it Department of Physics, 
Robeson Hall, Virginia Tech,\\ Blacksburg, VA 24061, U.S.A.}}

\fnote{}{cwei@vt.edu}
\fnote{}{jamesgray@vt.edu}

\end{center}

\begin{abstract}
\noindent 

	We discuss the extent to which numerical techniques for computing approximations to Ricci-flat metrics can be used to investigate hierarchies of curvature scales on Calabi-Yau manifolds. Control of such hierarchies is integral to the validity of curvature expansions in string effective theories. Nevertheless, for seemingly generic points in moduli space it can be difficult to analytically determine if there might be a highly curved region localized somewhere on the Calabi-Yau manifold. We show that numerical techniques are rather efficient at deciding this issue.

\end{abstract}

\thispagestyle{empty}
\setcounter{page}{0}
\newpage

\tableofcontents

\section{Introduction}

Ricci-flat metrics on Calabi-Yau manifolds are central objects in the study of compactifications in many string theoretic settings \cite{Green:1987mn,Polchinski:1998rr}. They frequently appear in a supergravity limit where curvatures are taken to be small with respect to the string scale and only a finite number of terms in an $\alpha'$ expansion have been kept. An assumption that is often made is that, if we are in a region of moduli space where the overall volume of the manifold is large and we are not near any singularities, then the values of curvature invariants on different points of the manifold will not become large and truncating the expansion at some finite order should be valid.

How do we know if this assumption is justified however? Could there be regions on the manifold where certain curvature invariants become large compared to the scale set by the overall volume in a manner that one would not naively expect? There are some choices of moduli for which such an expansion clearly breaks down. For example, for Calabi-Yau manifolds constructed as complete intersections in some simple ambient space, it is often easy to compute loci in moduli space, such as a conifold locus \cite{Candelas:1989ug,Candelas:1989js,Candelas:1990pi,Candelas:1990rm,Aspinwall:1993nu,Strominger:1995cz}, where the variety becomes non-transverse. Near to such singular loci higher order curvature invariants will not be controlled by the scale set by the overall volume and one would expect the supergravity approximation to break down. Similar comments could be made in regions of parameter space where certain cycles become small but non-vanishing.

The general question that may arise, however, is somewhat more difficult to answer. Say that we have some description of a Calabi-Yau manifold as an algebraic variety with a specific choice of coefficients in the defining polynomial (perhaps with these having been determined by some moduli stabilization mechanism). How do we decide if the approximations made in truncating the $\alpha'$ expansion are valid in such a case? It is not clear, for example, if there could be regimes of moduli space corresponding to Calabi-Yau manifolds exhibiting regions of very high curvature which are nevertheless not near to any singularity.  One might suspect that if the coefficients appearing in a defining relation are very large or very small that one might have an issue. More precisely, due to the overall scaling available in such defining relations, one might be concerned if large hierarchies in the sizes of coefficients that appeared were present. But how does one know for sure and what does one say if the coefficients are simply generic looking, seemingly innocuous, values? In this paper we wish to show that modern numerical methods for determining Ricci-flat metrics on Calabi-Yau manifolds \cite{Donaldson,Headrick:2005ch,Douglas:2006hz,Douglas:2006rr,Braun:2007sn,Braun:2008jp,Headrick:2009jz,Anderson:2010ke,Anderson:2011ed,Ashmore:2019wzb} are a practical and efficient tool for deciding such questions in many cases. 

The structure of the rest of this paper is as follows. We will begin, in Section \ref{numerics} with a review of some modern numerical techniques for finding approximations to Ricci-flat metrics on Calabi-Yau manifolds. We will also introduce some details of the examples we will be considering in the rest of the paper. In Section \ref{highcurv} we present our procedure for studying high curvature invariants numerically and apply these methods explicitly to a number of concrete examples. We demonstrate that we can reproduce expected hierarchies of curvature scales, associated to the presence of singularities, and describe how these methods can be used to address the issues described in this introduction. In Section \ref{conc} we briefly conclude and discuss possible future directions of research.

\section{Numerical Computation} \label{numerics}

In this section we briefly review some of the existing methods for obtaining numerical approximations to Ricci-flat metrics on Calabi-Yau manifolds \cite{Donaldson,Headrick:2005ch,Douglas:2006hz,Douglas:2006rr,Braun:2007sn,Braun:2008jp,Headrick:2009jz,Anderson:2010ke,Anderson:2011ed,Ashmore:2019wzb}. This is an evolving field with new techniques still being developed, notably recently including methods from Machine Learning \cite{Ashmore:2019wzb} (related work on Machine Learning and more general metrics with $SU(3)$ structure is currently underway \cite{ustoappear}). In this section, however, we will focus on the approach of \cite{Headrick:2009jz} which is the methodology which will be employed in this paper.

Most of the techniques for numerically computing Ricci-flat metrics on a Calabi-Yau $n$-fold $X$ begin with an ansatz for the K\"ahler potential on that manifold. Following \cite{Tian}, one can construct an ansatz of the following form.
\begin{eqnarray} \label{ansatz}
K_{h,k} = \frac{1}{k \pi} \ln \left(\bar{s}_{\bar{\beta}} h^{\bar{\beta} \alpha} s_{\alpha}\right)
\end{eqnarray}
In this expression the $s_{\alpha} \in H^0(X,{\cal L}^k)$ are a basis of global holomorphic sections for the $k$'th power of some ample holomorphic line bundle ${\cal L}$ over $X$. The ansatz (\ref{ansatz}) is labeled by two quantities, the $k$ of the previous sentence and the $h$ parameters, which we wish to adjust to make the associated metric as close as possible to Ricci-flat. K\"ahler potentials of the form (\ref{ansatz}) can be thought of as a generalization of the Fubini-Study case, where the powers of the variables involved have been increased and global sections of line bundles have replaced polynomials in coordinates of $\mathbb{P}^n$ in order to obtain a set of functions that are independent on $X$. As $k$ is increased there will be more and more global sections to ${\cal L}^k$ and thus more freedom in the ansatz, as controlled by the number of $h$ coefficients, to try and adjust to get close to a Ricci-flat metric. Indeed, it has been shown \cite{Donaldson,Donaldson2}, that in the limit as $k\to \infty$ a choice of the $h$ parameters exists where we approach precisely the Ricci-flat situation of interest.

Given an ansatz for the K\"ahler potential on a Calabi-Yau manifold of the form discussed in the last paragraph, the next step is to introduce a procedure for adjusting the parameters that appear in the ansatz to some optimal value where the associated metric is as close as possible to Ricci-flat. Here there are a couple of options available in the literature, including implementations of Donaldsons original proposal of the iteration of a ``T-operator" \cite{Donaldson,Donaldson2,Douglas:2006hz,Douglas:2006rr,Braun:2007sn,Braun:2008jp,Anderson:2010ke,Anderson:2011ed,Ashmore:2019wzb}. In this paper, however, we will use the proposal of Headrick and Nassar \cite{Headrick:2009jz}, wherein the coefficients in the ansatz (\ref{ansatz}) are optimized using the minimization of an appropriate functional.

Let us denote the no-where vanishing holomorphic $(n,0)$ form on $X$ by $\Omega$ and the K\"ahler form on the same space by $\omega$.  One can then define the following top forms.
\begin{eqnarray}
\mu_{\Omega} = (-i)^n \Omega \wedge \overline{\Omega} \;\;\; \textnormal{and} \;\;\; \mu_{\omega} = \frac{\omega^n}{n!}
\end{eqnarray}
Given that both of these objects are top forms, their ratio yields a function which will be denoted by $\eta$. We can also define the average of this function over $X$, denoted by $\left<\eta\right>$, which is a number.
\begin{eqnarray}
\eta = \frac{\mu_{\omega}}{\mu_{\Omega}} \;\; ,\;\;\left<\eta\right>=\frac{ \int_X \eta \,  \mu_{\Omega} }{\int_X \mu_{\Omega}}
\end{eqnarray} 
The functional that we will employ in this paper can then be defined in terms of these quantities \cite{Headrick:2009jz}.
\begin{eqnarray} \label{functional}
E(\omega) = \int_X (\eta -\left<\eta\right>)^2 \mu_{\Omega}
\end{eqnarray}
The idea is that, in many examples of interest, expressions for $\Omega$ are known analytically. One can then use the ansatz (\ref{ansatz}) to compute $\omega$ and thus $\eta$ and $E(\omega)$. From there, one can perform a minimization on $E$, to find the choice of the parameters $h$ that make that quantity as small as possible. In the limit $k\to \infty$ there is enough freedom in the K\"ahler potential ansatz to set $E$ to zero. Given the positive definite nature of the integrand in (\ref{functional}) this is a global minimum of the functional. At this minimum we have $\eta=\left<\eta\right>$ and thus $\omega^n = (-i)^n n!\left<\eta\right> \Omega \wedge \overline{\Omega}$. This is precisely the Monge-Ampere equation that should be solved in order to obtain a Ricci-flat metric. For any finite $k$ we would not expect that the ansatz (\ref{ansatz}) embodies the flexibility necessary to achieve an exact vanishing of (\ref{functional}). Instead we obtain an approximation to the Ricci-flat K\"ahler potential at each finite $k$, called the optimal metric \cite{Headrick:2009jz}, and simply increase the value of $k$ until a desired precision is reached.

At this stage it is useful to mention that, as noted in \cite{Headrick:2009jz}, one might expect an ansatz of the form (\ref{ansatz}) to be most useful for Calabi-Yau manifolds that are smooth and which do not exhibit a hierarchy of scales of curvature. For example, manifolds which are ``close" in moduli space to a nodal singularity will have highly curved regions in them and it is not clear that an ansatz of this form will be useful for any finite $k$ corresponding to a basis $s_{\alpha}$ of computationally manageable size. One of the burdens of this paper will be to show that (\ref{ansatz}) is indeed practically useful in mapping out where in moduli space such large curvature regions are occuring and that the above concerns do not prevent us from achieving useful results.

\subsection{Examples}\label{revieweg}
	
In this section we continue our review of the relevant pieces of \cite{Headrick:2009jz}, focussing on the setup for the particular examples we will be considering. For simplicity, in this paper we will discuss Calabi-Yau $n$-folds $X$ that are constructed as simple hypersurfaces in projective space. It is certainly possible to consider examples which are more complex in their description by utilizing the same methods that we are describing here. Restricting our attention to such well known examples, however, allows us to demonstrate that hierarchies of curvature scales can be located in moduli space in cases where there are known results to be compared to. There are also some specific technical advantages to this choice that we will mention at the end of this section. 

We consider an $n$-fold which is described as the vanishing locus of a degree $n+2$ polynomial, $p$, in $\mathbb{P}^{n+1}$.
\begin{equation} \label{CY}
 X=\left[
 \begin{array}{c|c}
 \mathbb{P}^{n+1}&n+2
 \end{array}
 \right]
\end{equation} 
For a threefold, for example, this gives us the famous degree $5$ quintic hypersurface in $\mathbb{P}^4$. Let us label the homogeneous coordinates of the ambient $\mathbb{P}^{n+1}$ by $\{z_{A} \},A=0,\ldots, n+1$. We will then break $\mathbb{P}^{n+1}$ up into $n+1$ polydiscs \cite{Anderson:2011ed}. More precisely we write
\begin{equation}
  \IP^{n+1} = \bigcup_{A=0}^N D_A
  ,\qquad
  D_A = 
  \Big\{
  [w_0:
  \cdots:w_{A-1}:
  w_{A+1}:\cdots:
  w_N]
  ~
  \Big|
  ~
  |w_j|\leq 1
  \Big\}\simeq D^{n+1}, 
\end{equation}
where the $w_{\alpha}=z_{\alpha}/z_A$ with $\alpha=0,\ldots, n+1$, omitting $\alpha=A$, are the local affine coordinates of the polydiscs $D_A$. In other words, given a specified set of homogeneous coordinates we divide all of the homogeneous coordinates by that with the largest magnitude in order to get the affine coordinates (and the number 1) for that point on the polydisc where it lives. Note that a generic point in $\mathbb{P}^{n+1}$ lies in only one polydisc, the overlaps being a set of measure zero. In addition the affine coordinates on each polydisc vary over a finite range, with the modulus being constrained to be less than or equal to one. Given these two facts we can see that this partitioning of the manifold is well suited to numerical integration techniques. Let $X_A$ be the restriction of $D_A$ to $X$. Obviously, these $X_A$'s then cover $X= \bigcup_{A=0}^N X_A$. On patch $X_A$, the coordinates $w_{\alpha}$ must obey the defining relation of the Calabi-Yau, expressed in suitable coordinates.
\begin{equation}
p(w_0,\ldots,w_{A-1},w_{A+1},\ldots,w_{N})=0
\end{equation}
Solving this equation for one variable, which we will denote as $w_{\delta}$, then leaves us with $n$ independent coordinates on each polydisc restricted to $X$, the correct number for a manifold of this dimension. The remaining independent coordinates on a given patch are then denoted by $\{w_i\},i=0,\ldots,N-1$.  

Over such a space a line bundle can be specified by its first Chern class, and thus the available ${\cal L}$ in constructing the ansatz (\ref{ansatz}) are simply given by ${\cal O}_X(k)$ where $c_1({\cal O}_X(k)) = k \omega$. Given that the $s_{\alpha}$ are then simply elements of $H^0(X,{\cal O}(k))$, they are given by a basis of the set of all degree $k$ polynomials in quotient ring associated to the defining relation $p$. A basis of this space is easily obtained in terms of the ambient space coordinates $w_A$. To obtain a metric from the K\"ahler potential (\ref{ansatz}), however, we must take derivatives with respect to coordinates on $X$ itself. This is most easily achieved by using the following manipulations \cite{Headrick:2009jz}.

Taking the $s_{\alpha}$ to be a basis of representatives of the equivalence classes of the coordinate ring of the right degree, we can, given a hermitian matrix $h$ write down the following metric on $\mathbb{P}^{n+1}$.
\begin{equation}
  \hat{g}_{\alpha\bar{\beta}}
  =\frac{1}{k\pi} \hat{\partial}_{\alpha}\hat{\partial}_{\bar{\beta}}  \ln \sum_{\alpha,\bar{\beta}=0}^{n_{k}-1} h^{\alpha\bar{\beta}}
  s_{\alpha}{\bar{s}}_{\bar{\beta}}
\end{equation}
There is then a simple relationship between this quantity and the metric we desire on $X$, $g_{i\bar{j}}=\partial_i \bar{\partial}_{\bar{j}} K_{h,k}$. This is given by the following.
\begin{equation} \label{metricX}
g_{i\bar{j}} = \hat{g}_{i\bar{j}} - \frac{p_i}{p_{\delta}} \hat{g}_{\delta \bar{j}} - \frac{\bar{p}_{\bar{j}}}{\bar{p}_{\bar{\delta}}} \hat{g}_{i\bar{\delta}} + \frac{p_i\bar{p}_{\bar{j}}}{|p_{\delta}|^2} \hat{g}_{\delta \bar{\delta}}
\end{equation}
In this expression the $p_i$ and $p_{\delta}$ are derivatives of $p$ with respect to the corresponding coordinates. A direct calculation using the standard expression for the holomorphic three form for such hypersurfaces \cite{AtiyahBottGarding,Candelas:1987kf},
\begin{eqnarray} \label{Omegaform}
\Omega = 
p_{\delta}^{-1}
\prod_{i=0}^{N-2}dw_i  \;,
\end{eqnarray}
 then shows that,
\begin{equation} \label{eta}
\eta= \frac{\mu_{\omega}}{\mu_{\Omega}} =\det(g_{i\bar{j}})|p_{\delta}|^2. 
\end{equation}  
This expression can then be used to evaluate $\eta$ for any given choice of the parameters $h$. 

\vspace{0.2cm}

To compute the functional (\ref{functional}) requires the additional step of performing an integral over the Calabi-Yau manifold $X$. This integral can be computed as the sum of integrals over the individual $X_A$ due to the fact that the overlap of these sets is of measure zero in the total space. 
\begin{equation} 
E=\sum_{A=0}^{N_s} E_A, \quad E_A = \int_{X_A} (\eta-1)^2 \mu_{\Omega}
\end{equation}

We will use the Monte Carlo method to compute each $E_A$ above: the integrals $E_A$ will be approximated with a sum over a randomly generated set of $N$ points $\{P_a \in X_A \}_{a=1}^{N}$.
\begin{equation}
E_A \approx \frac{V_A}{N}\sum_{a=1}^N (\eta(P_a) -1)^2
\end{equation}
Here $V_A= \int_{X_A} \mu_{\Omega}$ is the coordinate volume of the $A$'th restricted polydisc. To reduce the error in this approximation, it is useful to sample points in a manner that is uniformly distributed according to the volume form $\mu_{\Omega}$ of $X$ (see for example \cite{Douglas:2006rr}). In other words, an open set $U \in X$, the number of sample points taken in that set should be proportional to $\int_U \mu_{\Omega}$. 

Writing out the explicit expressions for $i^n\mu_{\Omega}$ given (\ref{Omegaform}) the following is found.
\begin{equation}
i^n\mu_{\Omega} = 
|p_{\delta}|^{-2}
\prod_{i=0}^{n}dw_i \wedge \prod_{\bar{j}=0}^{n} d\bar{w}_{\bar{j}}
\end{equation}
This can be rewritten in terms of ambient space quantities in the following manner.
\begin{equation}
i^n\mu_{\Omega} = \delta^2(p) \prod_{\alpha=0}^{n+1} dw_{\alpha} \wedge \prod_{\bar{\beta}=0}^{n+1} d\bar{w}_{\bar{\beta}}
\end{equation}
We can approximate this by spreading out the delta function with a small parameter $\epsilon$ as follows,
\begin{equation} \label{measureA}
i^n\mu_{\Omega} \approx \frac{1}{\epsilon^2}\Theta(\epsilon - |p|) \prod_{\alpha=0}^{N-1} dw_{\alpha} \wedge \prod_{\bar{\beta}=0}^{N-1} d\bar{w}_{\bar{\beta}} \;,
\end{equation}
so that the volume form now has support on a slab of thickness $\epsilon$ about $X$. Given (\ref{measureA}) one can simply do the following to obtain our set of points $\{P_a \in X_A \}_{a=1}^{N}$. First generate a sample set of $n_s$ points on $\mathbb{P}^{n+2}$ evenly distributed in the coordinate measure. Second, discard any points for which $|p| > \epsilon$. Finally, for the points that remain, project them onto the Calabi-Yau manifold $X$ using the Fubini-Study metric on $\mathbb{P}^{n+1}$. In this manner, we obtain a set of points distributed on $X$ in a good approximation to the measure given by $\mu_{\Omega}$.

Note that one might think that a simple sampling of this form would not be efficient when using such techniques to detect regions of moduli space where curvature hierarchies are occurring. In particular, one might think that some kind of adaptive mesh approach to point selection, including more points in the sample in regions of high curvature, might be required \cite{Anderson:2011ed}. While we have implemented such point selection strategies in a manner that works for general Calabi-Yau manifolds embedded in simple ambient spaces, in this paper we will show that the above simple procedure is in fact adequate for the application being pursued in this work.

\vspace{0.2cm}

With all of the above pieces in place, we can compute the relevant integrals required to obtain the energy functional (\ref{functional}) as a function of the parameters $h$ for fixed $k$. A Levenberg-Marquardt minimization procedure (for example) can then be run to find the values of the $h$'s that correspond to the optimal metric \cite{Headrick:2009jz}. This procedure can be repeated with increasing $k$ until a metric with the desired degree of accuracy is found (for which the optimized value of the energy functional (\ref{functional}) is sufficiently small for example).

\begin{figure}[!ht] \centering
\subfigure{\includegraphics[width=0.428\textwidth]{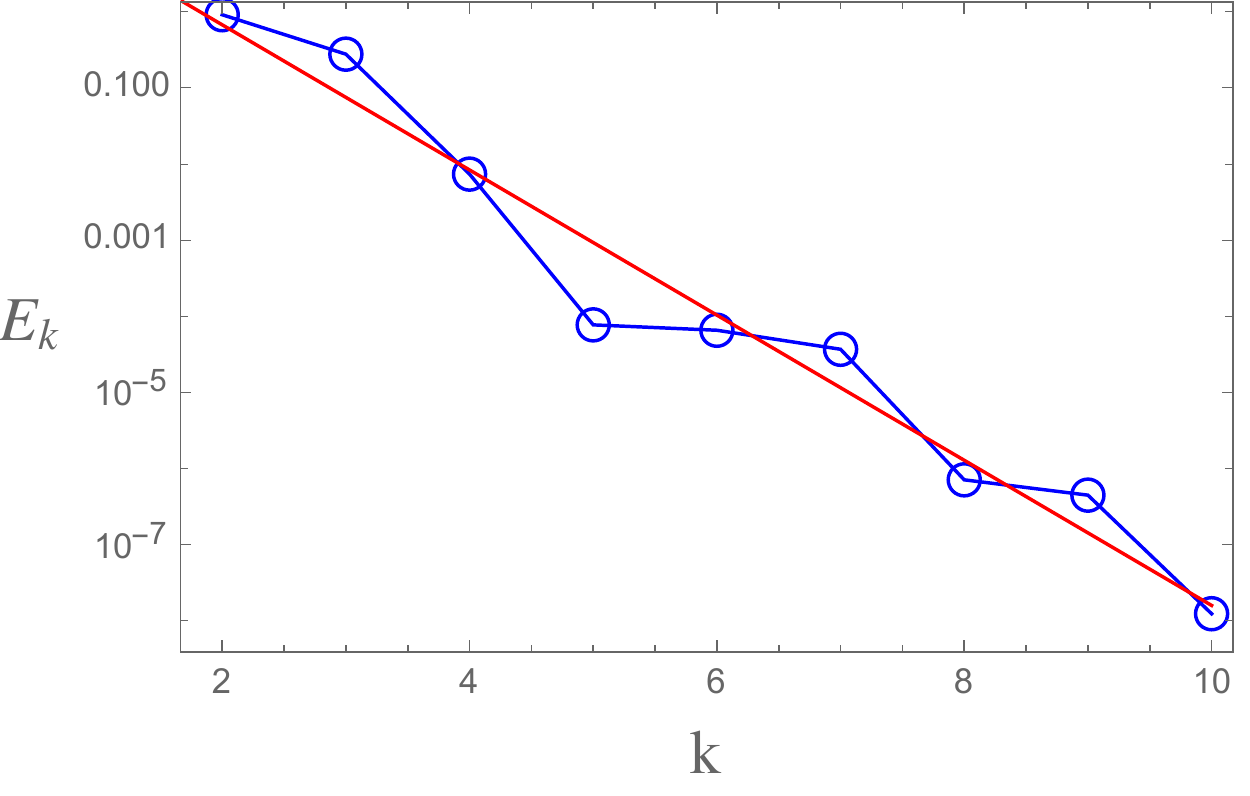}}
\subfigure{\includegraphics[width=0.428\textwidth]{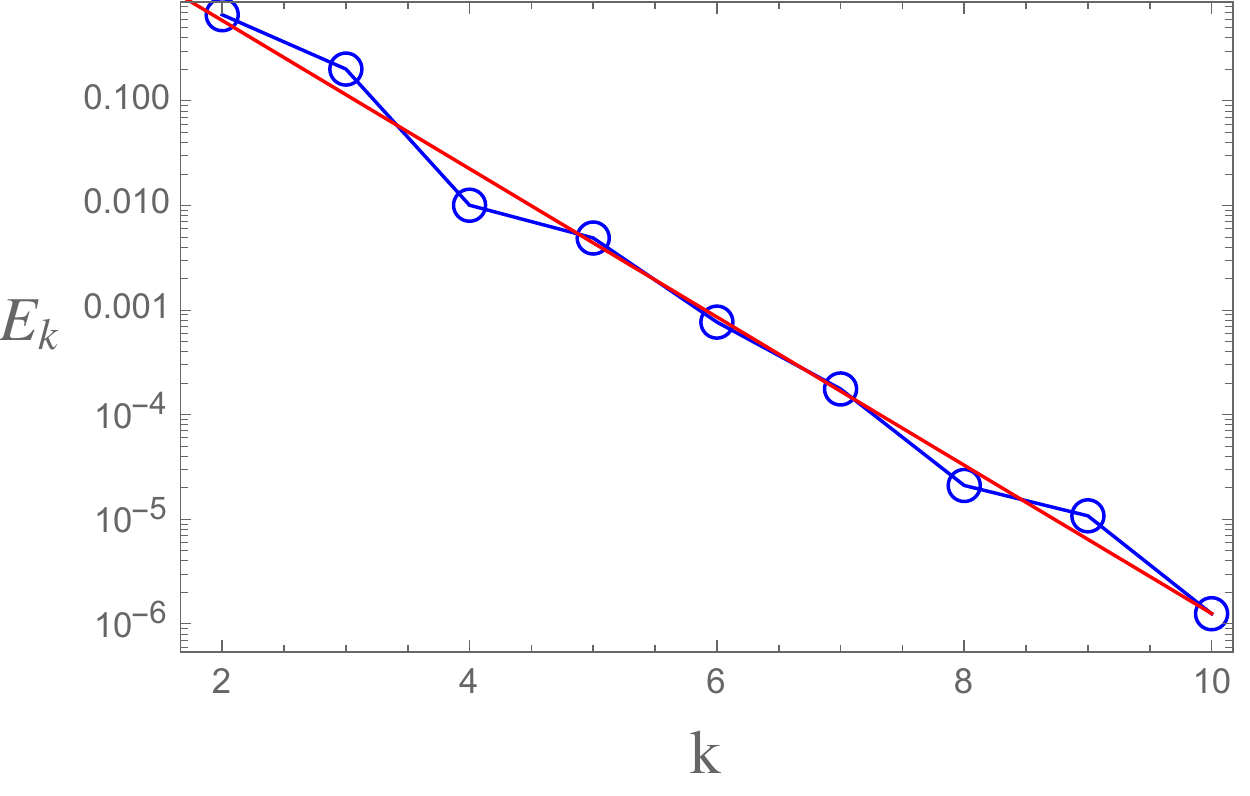}}
\caption{The error measure $E_k$ is simply the value of the functional (\ref{functional}) computed for the optimized metric at a given k for the Fermat quartic (left) and quintic (left). We see that the error measure approaches zero exponentially.}
\label{fig1}
\end{figure}

\vspace{0.3cm}
Using such methods it is possible to obtain Ricci-flat metrics on Calabi-Yau manifolds of various types to a high degree of accuracy. In Figure \ref{fig1} we illustrate this, in the case of the Fermat quartic and quintic Calabi-Yau two and three-folds respectively, by plotting the value of the functional (\ref{functional}), as a measure of the deviation from Ricci-flatness, as a function of $k$. We will not show such plots for all of the examples presented in this paper as the all look extremely similar to Figure \ref{fig1}, with exponential convergence of the energy functional to zero to a high degree of accuracy. For many of the examples considered in this paper, the freely available code found here \cite{Headrick:code} could be used to obtain the desired results. We have written our own code as a complement to this work however, which has been especially useful in studying cases not covered by \cite{Headrick:code} such as the two parameter example of Section \ref{2param}, or in considering the effectiveness of adaptive mesh style point selection techniques \cite{Anderson:2011ed}.

\vspace{0.2cm}

It should be noted that, practically, there is a serious issue that arises in implementing a procedure such as that described in this section. As $k$ is increased the dimension of $H^0(X,{\cal O}(k))$ increases rather rapidly. Given that $h$ in (\ref{ansatz}) has two indices running over this range, the number of parameters that the minimization must be performed with respect to increases even more quickly. In much of the current work on this subject this issue is dealt with by the imposition of a discrete symmetries. One starts with defining relations that preserve a large discrete symmetry group $\Gamma$ (which can exhibit fixed points on $X$) and only keep contributions to the ansatz (\ref{ansatz}) which are invariant under these symmetry actions. For example, for Fermat type defining relations for manifolds of the form (\ref{CY}) one could only keep those contributions which are invariant under the permutation of all ambient space coordinates and under phase changes of the coordinates by $n+2$ roots of unity. Such applications of symmetries dramatically reduce the number of contributions that need be considered at each $k$ and make the proliferation of parameters more manageable. In this paper we will impose all such symmetries which are admitted by the defining relations under investigation. Clearly, in the future development of such numeric techniques, it would be highly desirable to have a better resolution to this issue that would make it easier to deal with defining relations admitting no such symmetries.  
  
\section{Finding High Curvature Regions}  \label{highcurv}

\vspace{0.2cm}

Once we have a numerical approximation to the Ricci-flat metric on a given Calabi-Yau manifold, we then wish to address the central question of this paper. Given a set of defining relations for a Calabi-Yau variety can we determine if the Ricci-flat metric exhibits curvature scales that are wildly different from that set by the overall volume? To do this we apply the following procedure.

\begin{itemize}
    \item Find a numerical approximation to the Ricci-flat metric with K\"ahler potential of the form (\ref{ansatz}), as described in the previous section. This metric should be computed for as high a value of $k$ as possible, and the error measures associated to the metric should be checked to determine its accuracy.
    \item Find a sample of $n_R$ points on the Calabi-Yau manifold. Check that the Ricci-curvature is approximately zero on each of these points.
    \item Choose a number of derivatives $n_d$ and compute all curvature invariants that can be formed at this order. We consider all curvature invariants rather than just those appearing in a given $\alpha'$ expansion in the interests of generality. In addition, a large curvature appearing in a specific invariant might be a cause for concern, even if that invariant is not the particular one that appears in a given theory. One would suspect that such structure might indicate that high curvatures will appear in the $\alpha'$ expansion at some order, even if it is somehow evaded at the level of the number of derivatives being considered. 
    \item Evaluate these curvature invariants on the $n_R$ sampled points and record the highest value they attain. Denote this maximal value by $R^{m}_{I\;\textnormal{max}}$ where $m=n_d/2$ denotes the power of curvature invariants under consideration and $I$ runs over the different possible invariants. The index $I$ will be dropped in the case of $m=2$ where there is only a single non-vanishing independent invariant.
    \item Compute the following dimensionless quantities.
    \begin{eqnarray} \label{taudef}
\tau^m_I = (R^{m}_{I\;\textnormal{max}})^{\frac{1}{2m}} V_{\omega}^{1/2n}
\end{eqnarray}
In this expression $V_{\omega}= \int_X \mu_{\omega}$. This is the dimensionless ratio of the mass scale set by the curvature invariant to the mass scale set by the overall volume of $X$ (i.e. the compactification scale). The quantities $R^{m}_{I\;\textnormal{max}}$ are, of course, a rather meaningless in isolation as the metric can be multiplied by an arbitrary overall scale while maintaining a Ricci-flat solution, adjusting the curvature invariants in a correlated manner. The dimensionless quantity encapsulates whether there is a region of anomalously high curvature compared to the scale set by the size of the manifold. Once more, the index $I$ will be dropped in the case of $m=2$ where there is only a single non-vanishing independent invariant.
\end{itemize}
We perform all of the above steps at some fixed $k$ which, following the discussion of the previous section, is high enough to give a good approximation to the metric. We then repeat the computations at a much higher $k$ value to check the results do not change. The small differences between the results obtained in this manner give us a measure of the error in our results due to the numerical nature of the metric being used. In order to highlight the errors coming from the accuracy of the numerical metrics that are being used, we have included error bars associated to this source of uncertainty in the plots of results that will be presented throughout this section. For a more detailed discussion of the accuracy of numerical metrics obtained with the methods used in this work, particularly as one nears curvature singularities in moduli space, see \cite{Headrick:2009jz}.

One might think that in choosing the sample points in the second step in the bulleted list above, an adaptive mesh type of selection technique should be used. While we have indeed implemented such a sampling method to ensure that we are not missing any structure, in the examples we have studied this has proven to be unnecessary. No further high curvature regions are seen in these manifolds upon utilizing such a technique beyond those already successfully detected by the point sampling method described in the previous subsection. As such, all of the results presented in this publication were obtained using this latter technique.

\vspace{0.2cm}

In order to demonstrate that this procedure can indeed isolate regimes in moduli space where the $\alpha'$ expansion is breaking down, we apply this methodology to some simple cases where regions of high curvature, associated to singularity structure, are well understood. In particular we show we can detect the appearance of high curvatures near both conifold points and large complex structure limits. Given these results we expect that this methodology can act as a useful probe in deciding such issues more generally.

\vspace{0.3cm}

\subsection{Examples 1: One Parameter Families}

We will study several Calabi-Yau manifolds in the family described by \eqref{CY}. For the purposes of illustrating the methodology being proposed in this paper, we simplify our numerical computations by working on Calabi-Yau manifolds with discrete symmetries of high degree. We will begin by studying one parameter families of manifolds, in varying dimension, whose defining relation is of the following form. 
\begin{equation}\label{oneparaS}
p_{n}(\psi)=\sum_{i=0}^{n+1} z_i^{n+2}-(n+2) \psi \prod_{i=0}^{n+1}z_i 
\end{equation}
 Given such a defining relation, the order of the available symmetry group $\Gamma$ is $|\Gamma|=2 (N+1)^{(N - 1)}(N+1)!$ for generic values of $\psi$. This symmetry is enhanced at the point in moduli space corresponding to the Fermat type defining relation, $\psi=0$, where $|\Gamma|=2(N+1)^N(N+1)!$ \cite{Headrick:2009jz}. 

\vspace{0.2cm}

\begin{figure}[!ht] \centering
\subfigure{\includegraphics[width=0.4\textwidth]{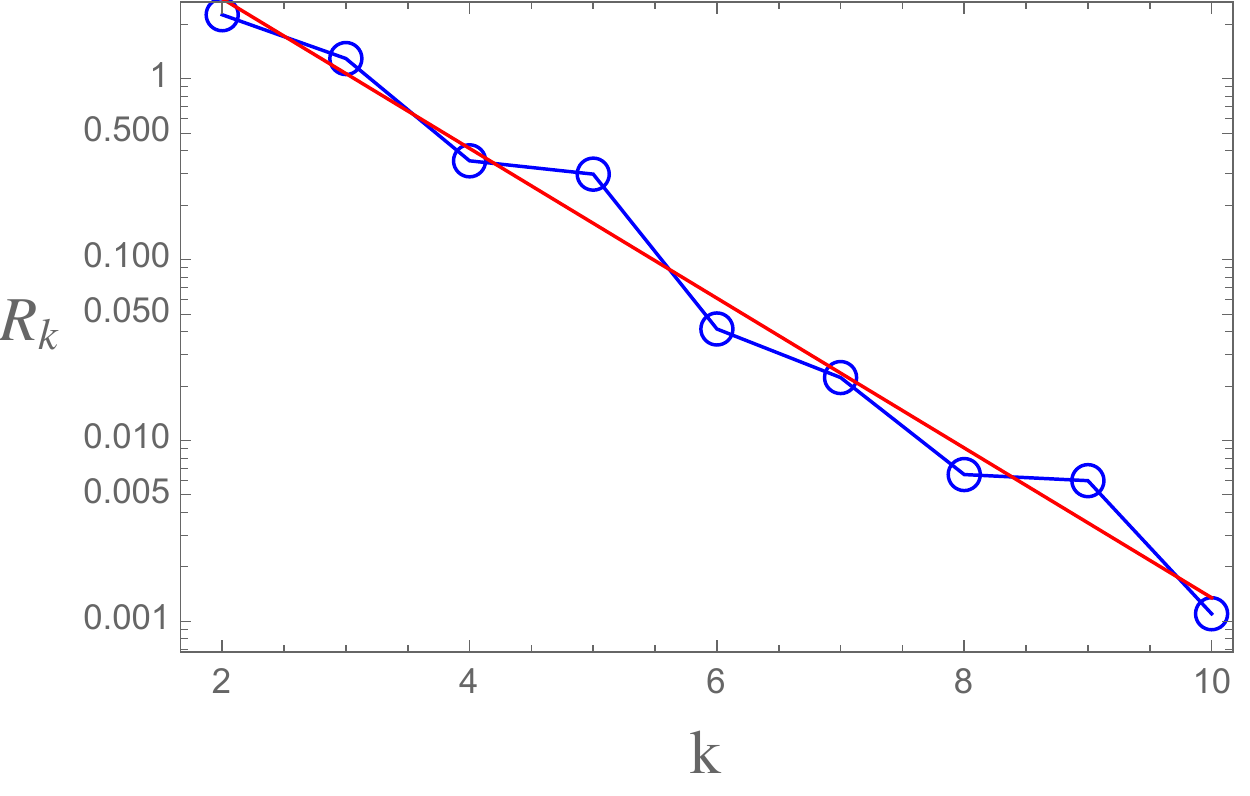}}
\subfigure{\includegraphics[width=0.41\textwidth]{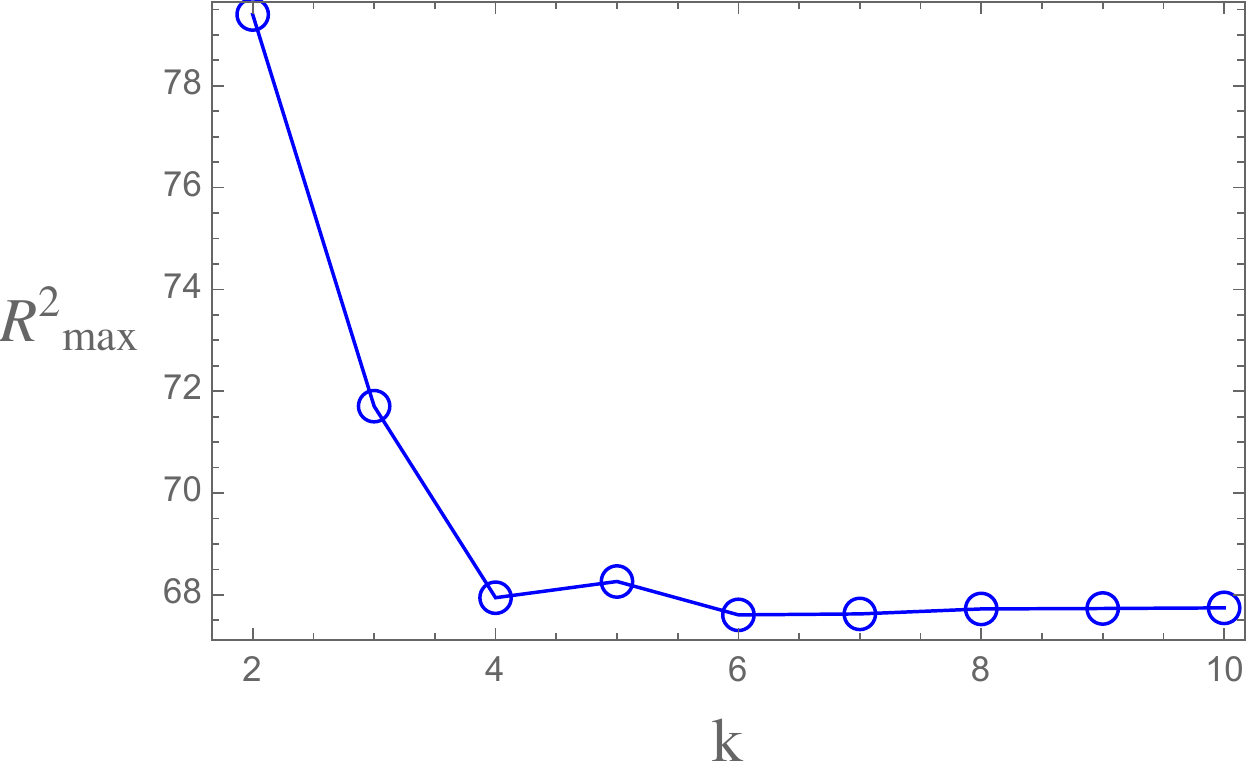}}
\caption{The maximum Ricci scalar (left) and magnitude of the curvature squared invariant (right) on the points sampled for the Fermat K3 in $\mathbb{P}^3$ as a function of $k$. The Ricci scalar approaches zero exponentially as $k$ is increased while the higher curvature invariant approaches a fixed value.}
\label{fig2}
\end{figure}

We begin by considering the one parameter family of quartics $p_2(\psi)$ in $\mathbb{P}^3$. In this example we utilize $n_s=10^4$ points and an accuracy of $\epsilon=0.02$ in terms of the parameters defined in Section \ref{revieweg}. We follow the procedure outlined at the start of this section for a variety of values of $\psi$ and for second order invariants in the curvature tensor. To demonstrate that we have good numerical control of the higher curvature invariants being computed, we present in Figure \ref{fig2} some plots for the example of the Fermat quartic $\psi=0$.
We see that the maximum value of the Ricci scalar on the $n_R=10^5$ sampled points rapidly approaches zero as we increase the accuracy of the numeric metric by increasing $k$. The maximum value of the higher order curvature invariant, by contrast, exponentially approaches a constant value - indicating that our accuracy is sufficient to trust the numerical result for this quantity.

\begin{figure}[!ht] \centering
\includegraphics[width=0.428\textwidth]{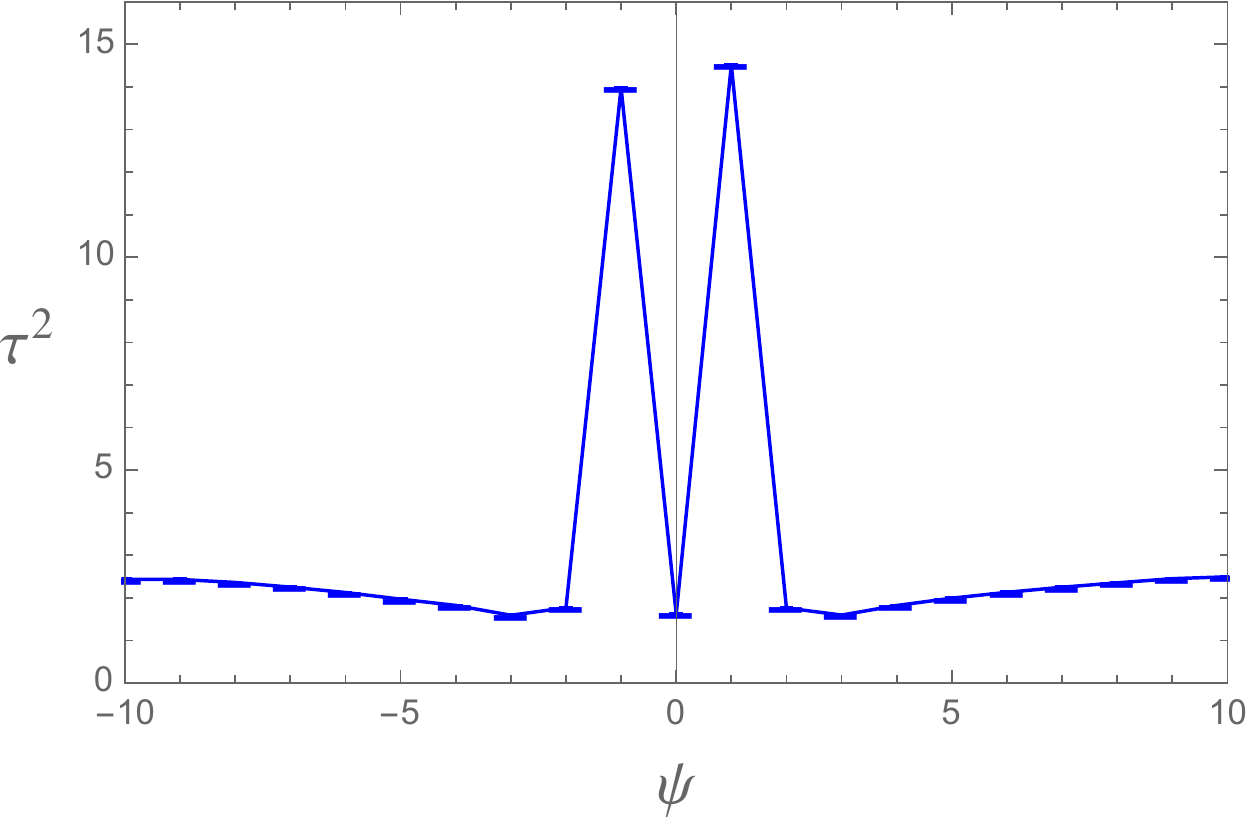}
\caption{The dimensionless measure of the maximum value of second order curvature invariants on the sampled points, $\tau^2$ from (\ref{taudef}), as a function of $\psi$ for the one parameter family of quartics in $\mathbb{P}^3$ (\ref{oneparaS}). The expected features in this plot, given the known location in moduli space of curvature singularities, are correctly reproduced.}
\label{fig3}
\end{figure}

\begin{figure}[!ht] \centering
\includegraphics[width=0.428\textwidth]{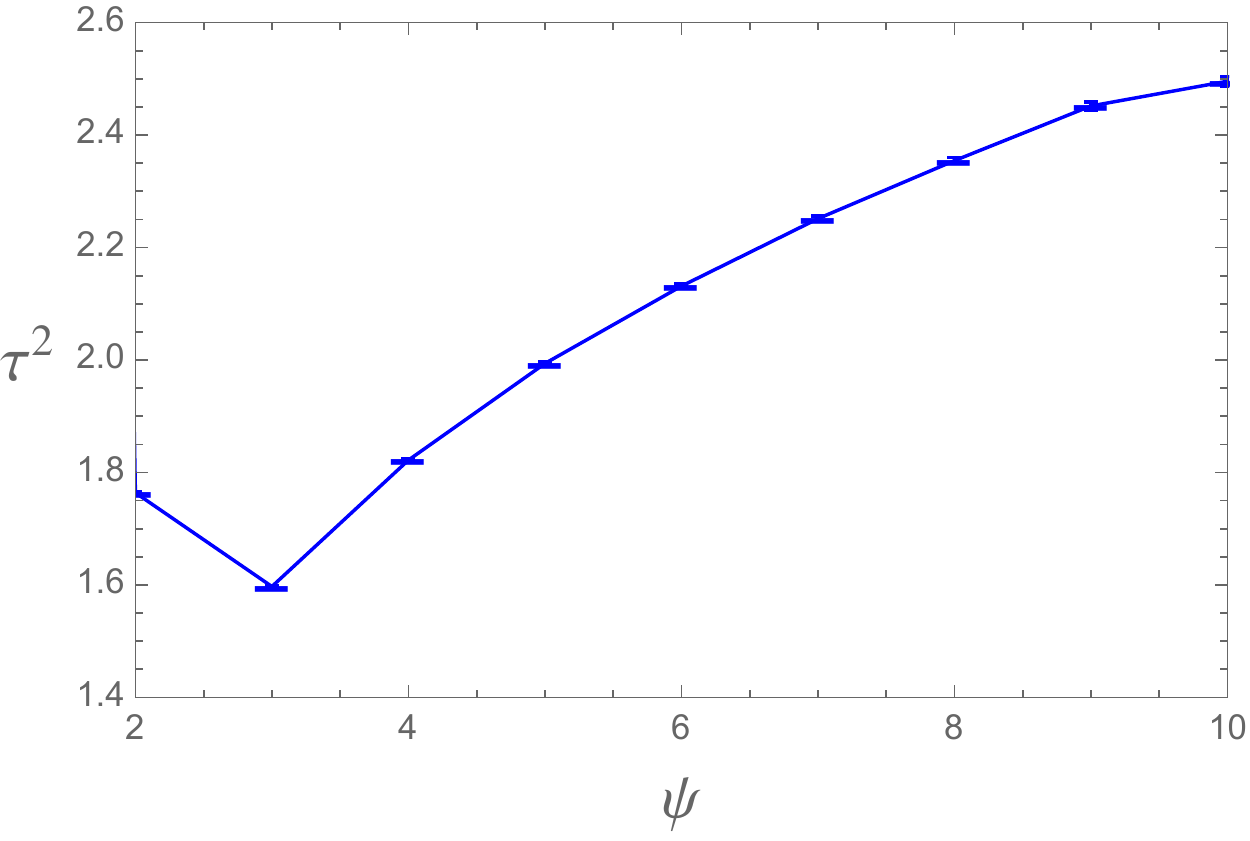}
\caption{The same data as presented in Figure \ref{fig3} with the large features associated to the conifold points omitted from the plot range. The tendency towards a hierarchy of scales between that set by the overall volume and that by highest value of the second order curvature invariant as the large complex structure limit is approached can clearly be seen.}
\label{fig4}
\end{figure}

Obtaining such results for a variety of values of $\psi$ and plotting the invariant $\tau^2$ to obtain a appropriately normalized measure of the size of the higher order curvature terms, we arrive at Figure \ref{fig3}. In compiling this plot we have computed initially with $k=5$ and ${\cal L}={\cal O}(1)$ and have checked the level of numerical error in our results by repeating the calculations at $k=10$.
The plot in Figure \ref{fig3} has several features that demonstrate that we can indeed isolate regimes of moduli space leading to higher curvature corrections correctly using numerical methods. First, the $K3$ of the form we are discussing has conifold singularities at $\psi=\pm 1$. The associated spikes in $\tau^2$, which is not infinite because a finite sampling of points will not land exactly on a singular point in the manifold, can clearly be seen in Figure \ref{fig3}. In addition to this obvious feature, one can also see that $\tau^2$ tends to increase as $|\psi|$ gets larger. This can be seen more clearly in a plot omitting the large features due to the conifold points, as in Figure \ref{fig4}. This corresponds to the increase in curvature scales appearing in the Ricci-flat metric as the manifold approaches the large complex structure limit.

\vspace{0.2cm}

Although the above analysis was for an algebraic $K3$ surface, very similar results can be obtained for threefolds and indeed higher dimensional varieties. Here we will content ourselves with presenting analogous results for the quintic Calabi-Yau threefold, using ${\cal L}={\cal O}(1)$, $n_s=10^4$, $n_R=10^5$ and $\epsilon=0.02$, in order to illustrate a new phenomenon that occurs in odd dimensions.

\begin{figure}[!ht] \centering
\subfigure{\includegraphics[width=0.4\textwidth]{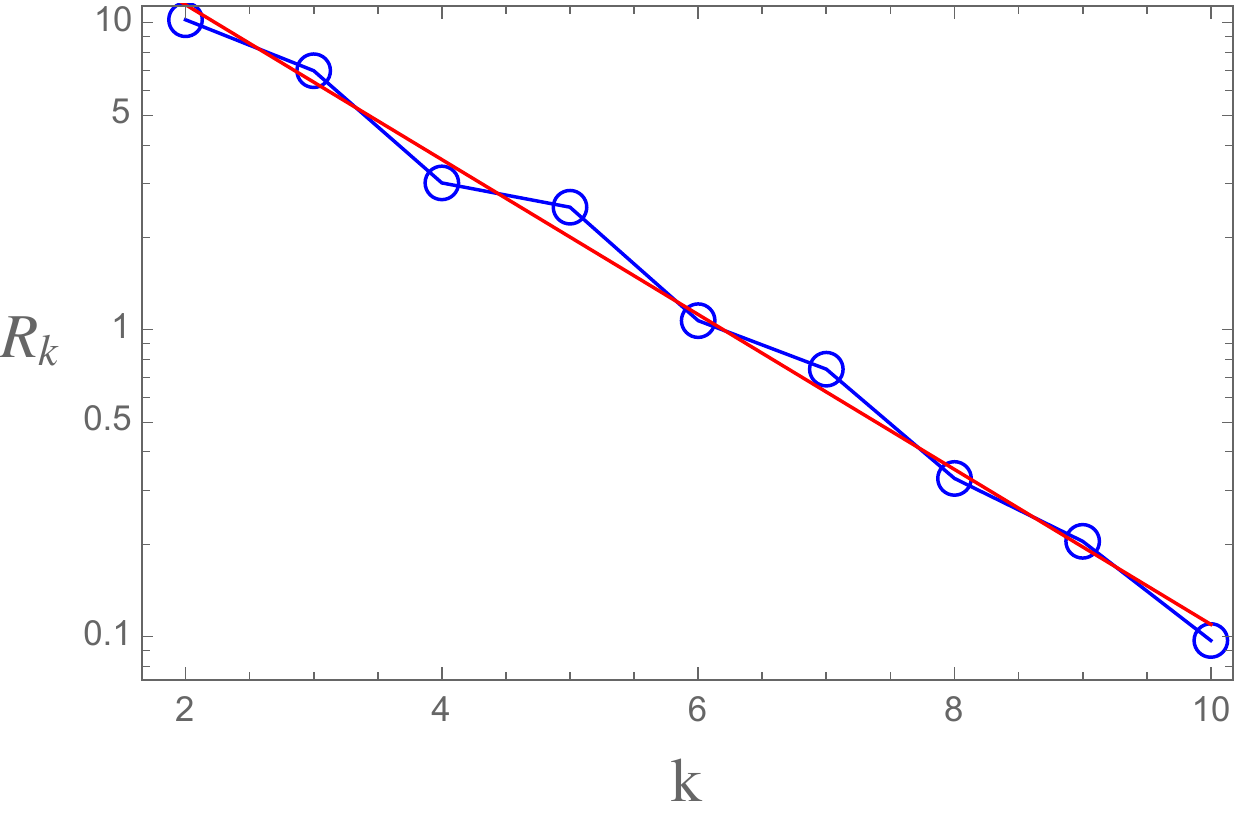}}
\subfigure{\includegraphics[width=0.428\textwidth]{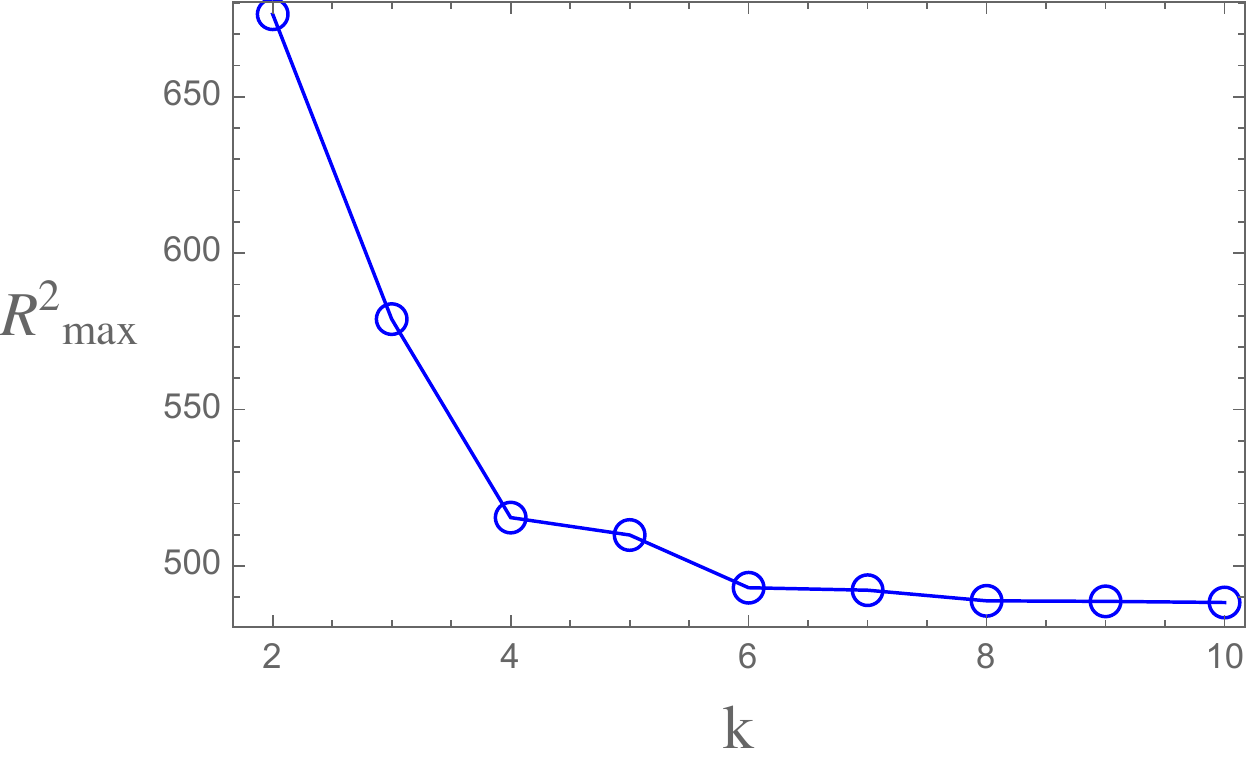}}
\caption{The maximum Ricci scalar (left) and magnitude of the curvature squared invariant (right) on the points sampled for the Fermat quintic threefold in $\mathbb{P}^4$ as a function of $k$. The Ricci scalar approaches zero exponentially as $k$ is increased while the higher curvature invariant approaches a fixed value.}
\label{fig5}
\end{figure}

\begin{figure}[!ht] \centering
\subfigure{\includegraphics[width=0.4\textwidth]{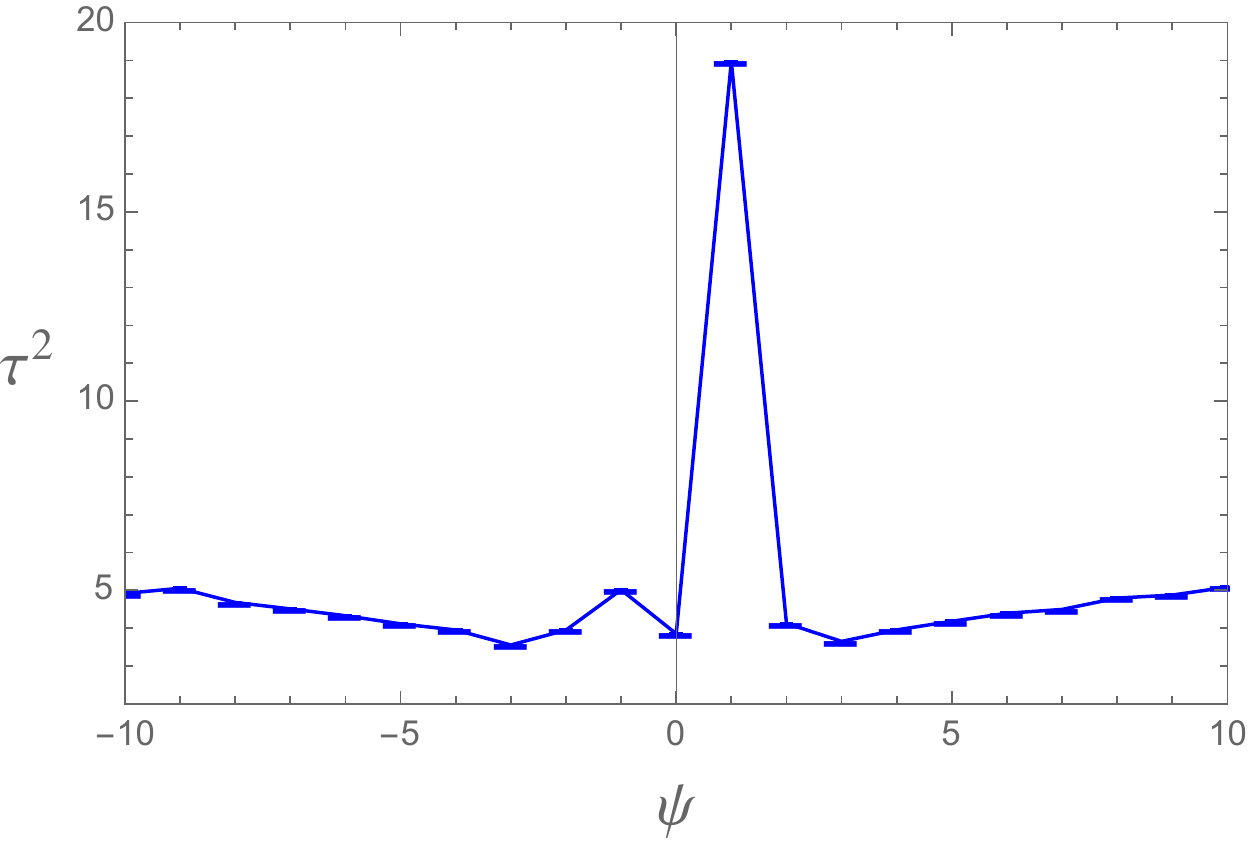}}
\subfigure{\includegraphics[width=0.428\textwidth]{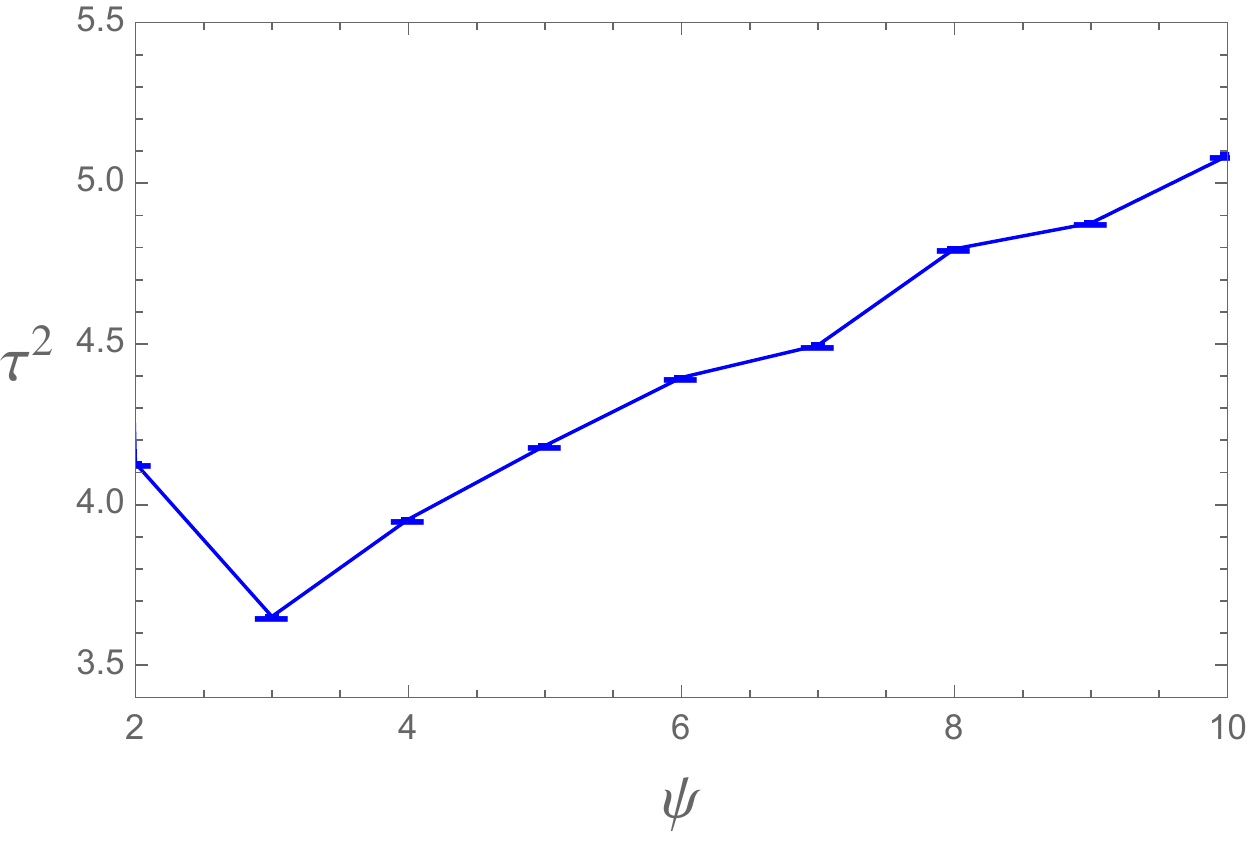}}
\caption{The measure of second order curvature invariants $\tau^2$, as a function of $\psi$ for the one parameter family of quintics in $\mathbb{P}^4$ (\ref{oneparaS}). The expected features can be seen in the left hand plot, given the known location in moduli space of curvature singularities, are correctly reproduced. An additional small feature is also seen at $\psi=-1$. The same data is presented in the right hand plot with the large features associated to the conifold points omitted from the plot range. The tendency of $\tau^2$ to increase as we approach the large complex structure limit can clearly be seen.}
\label{fig6}
\end{figure}

 In Figures \ref{fig5} and \ref{fig6} we present the analogous plots to those presented above in the $K3$ case for quintic Calabi-Yau threefolds in $\mathbb{P}^4$ of the form (\ref{oneparaS}).
  It is important to note that there is only one conifold point in moduli space in these quintic examples, at $\psi=1$. Nevertheless, in Figure \ref{fig6} a definite, albeit smaller, peak can be seen in the plot of $\tau^2$ against complex structure at $\psi=-1$. The variety is completely smooth at this point and is not ``near to a singularity" in any normal sense.  To check that this is a real effect one can repeat the computation, at both $\psi=-1$ and $\psi=1$, with an increased number of points $n_R=10^7$. The result at $\psi=1$ changes dramatically when we do this with $\tau^2$ changing from $11.2$ to $19.0$ (recall the error bars in the plots here represent the uncertainty due to the numerical nature of the metric and do not include errors due to point sampling in computing curvatures - which are negligible except at the singular points in moduli space). This is because as we compute the curvature on more and more points, we will randomly pick out curvatures that are closer and closer to the singularity, where $\tau^2$ diverges. The result at $\psi=-1$, however, hardly changes at all in changing $n_R$ by two orders of magnitude, with $\tau^2$ simply varying from $4.96$ to $5.01$. In this case we already have a good approximation to the maximum value of the finite range of values of $\tau^2$ at various points in this smooth manifold.
 
Although it does not lead to a particularly large hierarchy of scales in this case, this is a feature of the type we were looking to find. The increase in $\tau^2$ at $\psi=-1$ is corresponds to a Ricci-flat metric which exhibits rather large variations in curvature compared to nearby metrics in complex structure moduli space but which is nevertheless not near to any singularity. It would be seemingly rather difficult to see analytically that the defining relation with $\psi=-1$ leads to such behavior. Nevertheless, the numerical methods being used here can show this clearly. In cases where such effects are more pronounced, knowledge of this type would be vital in understanding where the $\alpha'$ expansions of string theories are valid. It is intriguing that the threefold case somehow `remembers' that the analogous $K3$ has a curvature singularity at this point in moduli space in this fashion. It would be interesting to have an analytic understanding of this phenomenon (see \cite{Headrick:2009jz} for a discussion of a potential explanation and related structure, albeit not directly linked to higher curvatures, at the point $\psi=-1$).  
 
 Note that, in addition to the features described above, it is just as important that elsewhere in the space of possible $\psi$ we are {\it not} seeing hierarchies of higher curvature scales. Confirming this result, while consistent with naive expectations, was one of the goals of this research.

\subsection{Examples 2: A Two Parameter Family} \label{2param}

In order to provide one final example which is not of the ``$\psi$ deformed Fermat" type we will consider a $K3$ surface embedded inside $\mathbb{P}^3$ with a defining relation depending upon two parameters of the following form.
\begin{equation} \label{twoparaS}
P_{2}(\psi_1,\psi_2)=z_0^4+z_1^4+\psi_1(z_2^4+z_3^4)+\psi_2z_0^2z_1^2
\end{equation}
In this case, the order of discrete symmetry that is available to simplify our computations is only $|\Gamma|=16$. This leads to a number of technical complications in completing the analysis of this example. For example, one can not use symmetry relationships between patches on the manifold to reduce the number of separate polydiscs that need to be considered. One also needs to consider more contributions to (\ref{ansatz}) at each given $k$. Nevertheless, it is straightforward enough to obtain accurate results in this case where we use ${\cal L}={\cal O}(1)$, $\epsilon=0.02$, $n_s=10^4$ and $n_R= 5\times 10^5$.

\begin{figure}[!ht] \centering
\subfigure{\includegraphics[width=0.4\textwidth]{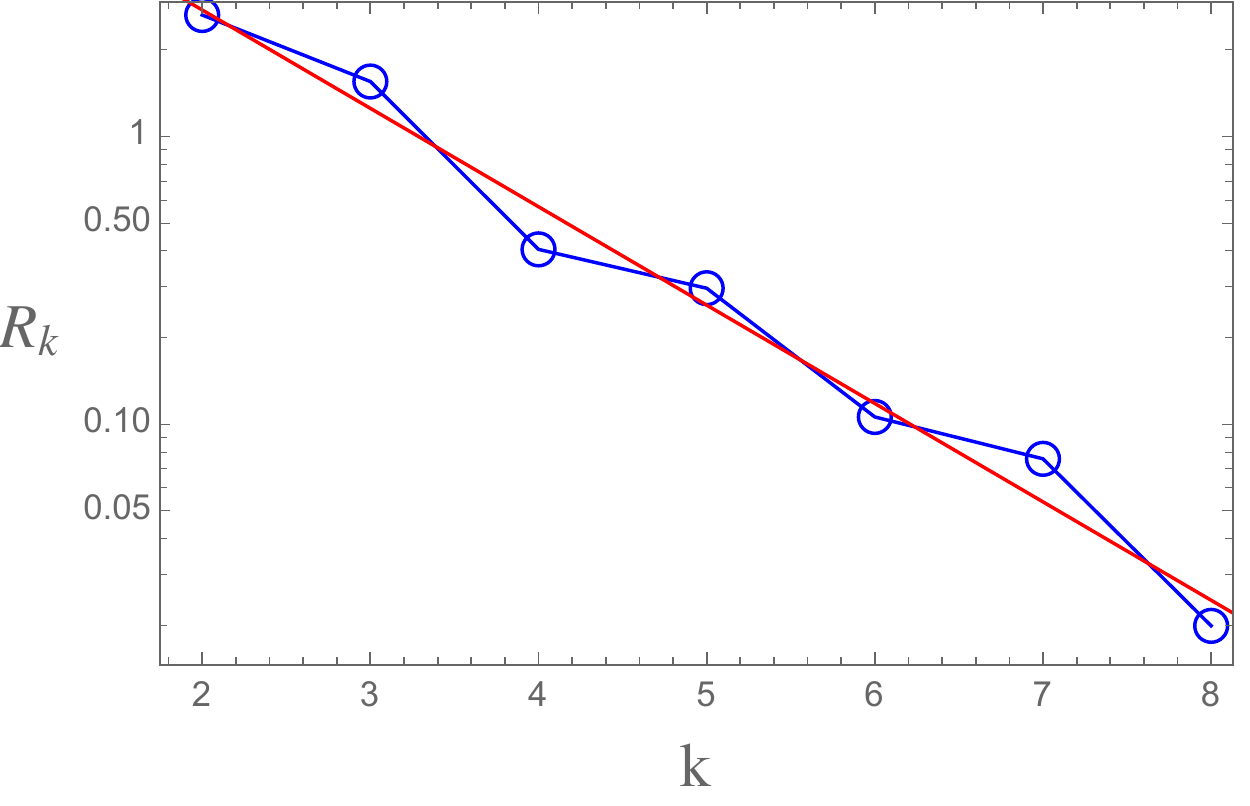}}
\subfigure{\includegraphics[width=0.428\textwidth]{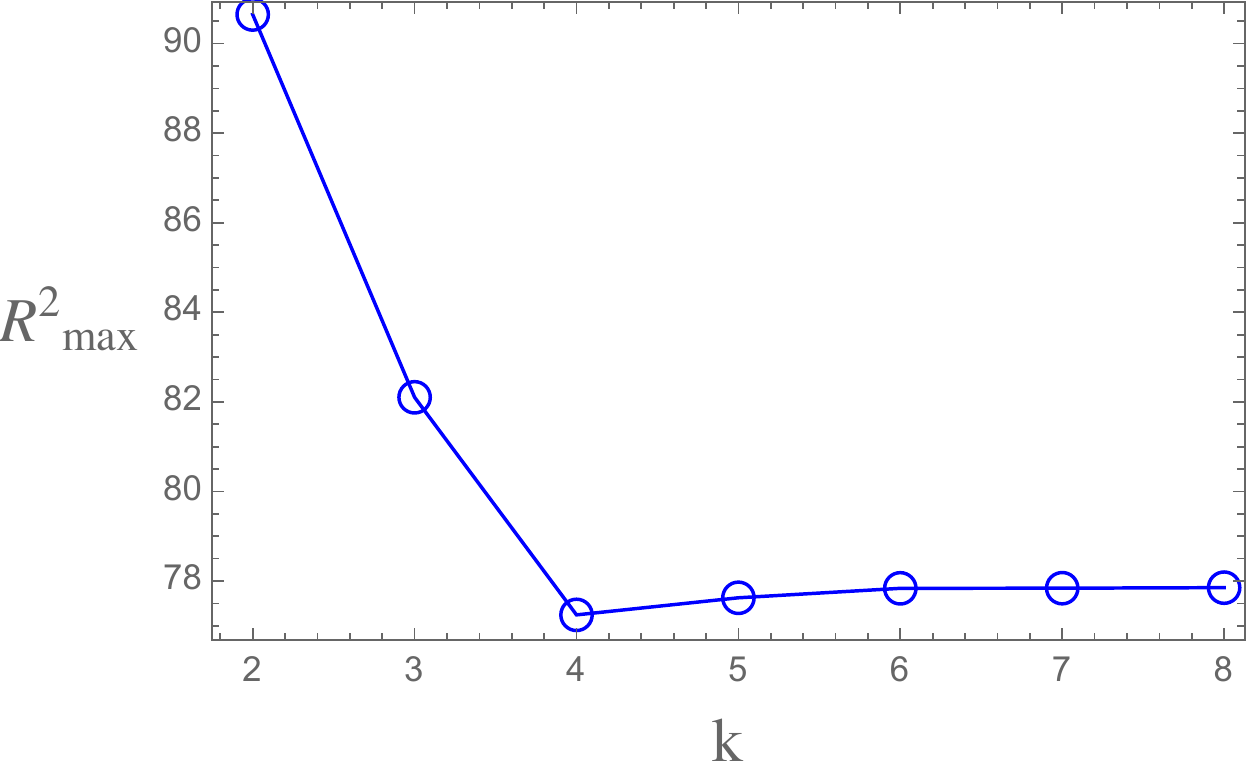}}
\caption{The maximum Ricci scalar (left) and magnitude of the curvature squared invariant (right) on the points sampled for the quartic in $\mathbb{P}^3$ given in (\ref{twoparaS}), with $\psi_1=1$ and $\psi_2=8$, as a function of $k$. The Ricci scalar approaches zero exponentially as $k$ is increased while the higher curvature invariant approaches a fixed value.}
\label{fig7}
\end{figure}

The reliability of the results in this case can be seen, for example, from the plots in Figure \ref{fig7}.
Here we present, for the example case of $\psi_1=1$ and $\psi_2=8$, that the maximum value of the Ricci scalar once more goes exponentially to zero with increasing $k$ while the maximum value of the higher curvature invariant approaches a constant value.

\begin{figure}[!ht] \centering
\includegraphics[width=0.428\textwidth]{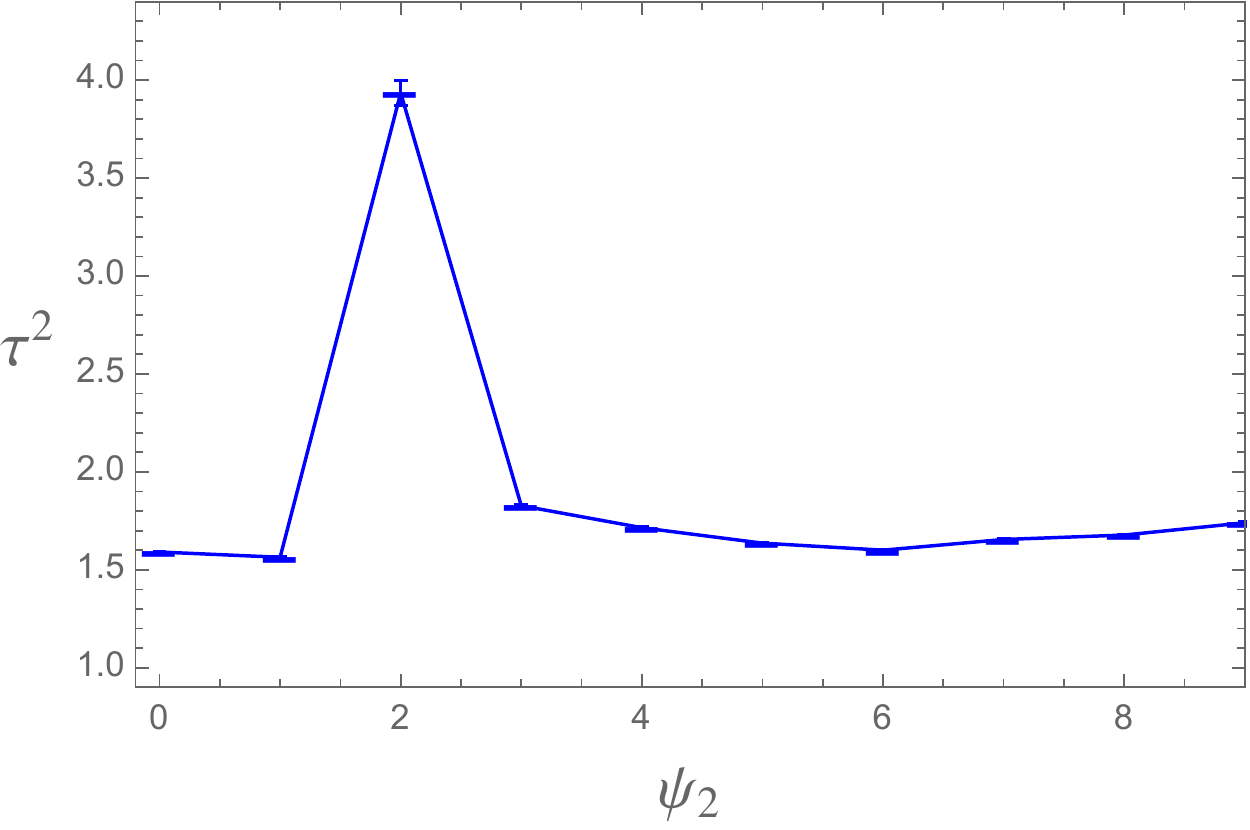}
\caption{The dimensionless measure of the maximum value of second order curvature invariants on the sampled points, $\tau^2$ from (\ref{taudef}), as a function of $\psi_2$ with $\psi_1=1$ for the two parameter family of quartics in $\mathbb{P}^3$ (\ref{oneparaS}). The expected features in this plot, given the known location in moduli space of curvature singularities, are correctly reproduced.}
\label{fig8}
\end{figure}

As with previous examples we can now explore the parameter space of the defining relation (\ref{twoparaS}) and show how the normalized measure of the maximum value of the second order curvature invariant $\tau^2$ varies with parameters. Since this example does not yield any surprises, we will content ourselves with showing a single plot, given in Figure \ref{fig8}, that illustrates that our numerical results reproduce the expected structure.
The defining relation (\ref{twoparaS}) exhibits singularities at $\psi_1=1\;,\;\psi_2=2$ and as $\psi_2 \to \infty$ at constant $\psi_1$. The singularity at finite parameters is clearly visible just as in previous examples and a steady increase in $\tau_2$ as $\psi_2$ approaches large values is also present as expected.

\section{Conclusions and Future Directions} \label{conc}

In this paper we have studied to what extent numerical methods can be utilized to detect hierarchies of curvature scales appearing in Ricci-flat metrics on Calabi-Yau manifolds at different locations in moduli space. These hierarchies concern a comparison of the scale set by the volume of the Calabi-Yau manifold to those determined by higher order curvature invariants. By illustrating that we can reproduce the expected behavior of such quantities as the system approaches singular points in complex structure moduli space, we have demonstrated that such techniques are rather effective in deciding this issue. Indeed, this is true even if simple point finding strategies for performing numerical integrations on the Calabi-Yau manifold are used: one does not necessarily need to adopt adaptive mesh procedures as one might suspect. 

In addition to reproducing expected structure, we have illustrated one feature that might be hard to find analytically but which can be isolated with numerical techniques. We have found a point in the moduli space of the quintic Calabi-Yau threefold wherein anomalously large variations of curvature are seen in the Ricci-flat metric when compared to nearby choices of complex structure. It is interesting to note that the analogous choice of complex structure in similar Calabi-Yau manifolds of even dimension would result in a curvature singularity, even though the threefold case being considered is itself perfectly smooth.

These techniques are a useful tool in deciding where in moduli space $\alpha'$ expansions, that are commonly used in constructing effective theories describing string compactifications, are valid. The utility of these methods is only likely to increase as advances are made in techniques for computing Ricci-flat metrics numerically.

\vspace{0.1cm}

Many future extensions of the type of work carried out here could be envisioned. One could use numerical approximations to the gauge connection in heterotic theories \cite{Douglas:2006hz,Anderson:2010ke,Anderson:2011ed}, for example, to study similar issues in the validity of $\alpha'$ expansions in those contexts. One could also study not just the complex structure dependence of hierarchies of curvature scales, but also their variation with K\"ahler moduli. In particular, in principle numerical methods could be used to delineate the boundaries of the K\"ahler cones of Calabi-Yau geometries: information which is often quite difficult to obtain (see \cite{Anderson:2017aux} for just one recent study where such considerations were the key limiting factor). However, the authors suspect that such a study would require an improvement in the currently known numerical techniques for finding approximations to Ricci-flat metrics. To divide the K\"ahler cone finely enough to detect the desired structure, one would have to consider polarizations including rather large numbers. This would lead to issues with computational complexity as the number of parameters appearing in the standard ansatz for the K\"ahler potential (\ref{ansatz}) would become large very quickly as one tried to obtain an accurate approximation to the metric.

\section*{Acknowledgments}

W.C. would like to thank M. Headrick for valuable discussions. The work of W.C. and J.G. is supported in part by NSF grant PHY-1720321. The authors would like to gratefully acknowledge the Simons Center for Geometry and Physics (and the semester long program, The Geometry and Physics of Hitchin Systems) for hospitality during the completion of this work.



\begin{thebibliography}{99}



\bibitem{Green:1987mn} 
  M.~B.~Green, J.~H.~Schwarz and E.~Witten,
  ``Superstring Theory. Vol. 2: Loop Amplitudes, Anomalies And Phenomenology,''
  Cambridge, Uk: Univ. Pr. ( 1987) 596 P. ( Cambridge Monographs On Mathematical Physics)
 
\bibitem{Polchinski:1998rr} 
  J.~Polchinski,
  ``String theory. Vol. 2: Superstring theory and beyond,''
  doi:10.1017/CBO9780511618123
 
 \bibitem{Candelas:1989ug} 
  P.~Candelas, P.~S.~Green and T.~Hubsch,
  ``Rolling Among Calabi-Yau Vacua,''
  Nucl.\ Phys.\ B {\bf 330}, 49 (1990).
  doi:10.1016/0550-3213(90)90302-T
 
 \bibitem{Candelas:1989js} 
  P.~Candelas and X.~C.~de la Ossa,
  ``Comments on Conifolds,''
  Nucl.\ Phys.\ B {\bf 342}, 246 (1990).
  doi:10.1016/0550-3213(90)90577-Z
 
 \bibitem{Candelas:1990pi} 
  P.~Candelas and X.~de la Ossa,
  ``Moduli Space of {Calabi-Yau} Manifolds,''
  Nucl.\ Phys.\ B {\bf 355}, 455 (1991).
  doi:10.1016/0550-3213(91)90122-E
  
  \bibitem{Candelas:1990rm} 
  P.~Candelas, X.~C.~De La Ossa, P.~S.~Green and L.~Parkes,
  ``A Pair of Calabi-Yau manifolds as an exactly soluble superconformal theory,''
  Nucl.\ Phys.\ B {\bf 359}, 21 (1991)
  [AMS/IP Stud.\ Adv.\ Math.\  {\bf 9}, 31 (1998)].
  doi:10.1016/0550-3213(91)90292-6
  
  \bibitem{Aspinwall:1993nu} 
  P.~S.~Aspinwall, B.~R.~Greene and D.~R.~Morrison,
  ``Calabi-Yau moduli space, mirror manifolds and space-time topology change in string theory,''
  Nucl.\ Phys.\ B {\bf 416}, 414 (1994)
  [AMS/IP Stud.\ Adv.\ Math.\  {\bf 1}, 213 (1996)]
  doi:10.1016/0550-3213(94)90321-2
  [hep-th/9309097].

 \bibitem{Strominger:1995cz} 
  A.~Strominger,
  ``Massless black holes and conifolds in string theory,''
  Nucl.\ Phys.\ B {\bf 451}, 96 (1995)
  doi:10.1016/0550-3213(95)00287-3
  [hep-th/9504090].
  
    
  \bibitem{Donaldson} 
S.~K.~Donaldson, 
``Some numerical results in complex differential geometry,"
Pure\ Appl.\ Math.\ Q.\ {\bf 5}, no. 2, part 1 571-618 (2009)
doi:10.4310/PAMQ.2009.v5.n2.a2
[arXiv:math/0512625].
  
  \bibitem{Headrick:2005ch} 
  M.~Headrick and T.~Wiseman,
  ``Numerical Ricci-flat metrics on K3,''
  Class.\ Quant.\ Grav.\  {\bf 22}, 4931 (2005)
  doi:10.1088/0264-9381/22/23/002
  [hep-th/0506129].
  
  \bibitem{Douglas:2006hz} 
  M.~R.~Douglas, R.~L.~Karp, S.~Lukic and R.~Reinbacher,
  ``Numerical solution to the hermitian Yang-Mills equation on the Fermat quintic,''
  JHEP {\bf 0712}, 083 (2007)
  doi:10.1088/1126-6708/2007/12/083
  [hep-th/0606261].
  
  \bibitem{Douglas:2006rr} 
  M.~R.~Douglas, R.~L.~Karp, S.~Lukic and R.~Reinbacher,
  ``Numerical Calabi-Yau metrics,''
  J.\ Math.\ Phys.\  {\bf 49}, 032302 (2008)
  doi:10.1063/1.2888403
  [hep-th/0612075].

     \bibitem{Braun:2007sn} 
  V.~Braun, T.~Brelidze, M.~R.~Douglas and B.~A.~Ovrut,
  ``Calabi-Yau Metrics for Quotients and Complete Intersections,''
  JHEP {\bf 0805}, 080 (2008)
  doi:10.1088/1126-6708/2008/05/080
  [arXiv:0712.3563 [hep-th]].
  
  \bibitem{Braun:2008jp} 
  V.~Braun, T.~Brelidze, M.~R.~Douglas and B.~A.~Ovrut,
  ``Eigenvalues and Eigenfunctions of the Scalar Laplace Operator on Calabi-Yau Manifolds,''
  JHEP {\bf 0807}, 120 (2008)
  doi:10.1088/1126-6708/2008/07/120
  [arXiv:0805.3689 [hep-th]].
  
  \bibitem{Headrick:2009jz} 
  M.~Headrick and A.~Nassar,
  ``Energy functionals for Calabi-Yau metrics,''
  Adv.\ Theor.\ Math.\ Phys.\  {\bf 17}, no. 5, 867 (2013)
  doi:10.4310/ATMP.2013.v17.n5.a1
  [arXiv:0908.2635 [hep-th]].

\bibitem{Anderson:2010ke} 
  L.~B.~Anderson, V.~Braun, R.~L.~Karp and B.~A.~Ovrut,
  ``Numerical Hermitian Yang-Mills Connections and Vector Bundle Stability in Heterotic Theories,''
  JHEP {\bf 1006}, 107 (2010)
  doi:10.1007/JHEP06(2010)107
  [arXiv:1004.4399 [hep-th]].
  
  \bibitem{Anderson:2011ed} 
  L.~B.~Anderson, V.~Braun and B.~A.~Ovrut,
  ``Numerical Hermitian Yang-Mills Connections and Kahler Cone Substructure,''
  JHEP {\bf 1201}, 014 (2012)
  doi:10.1007/JHEP01(2012)014
  [arXiv:1103.3041 [hep-th]].
 
  \bibitem{Ashmore:2019wzb} 
  A.~Ashmore, Y.~H.~He and B.~A.~Ovrut,
  ``Machine learning Calabi-Yau metrics,''
  arXiv:1910.08605 [hep-th].
  
  \bibitem{ustoappear}
  L.~B.~Anderson, J.~Gray, S.~Krippendorf, N.~Raghuram and F.~Ruehle,
  ``Machine learning SU(3) structure,"
  To appear.
  
  \bibitem{Tian}
G.~Tian, 
``On a set of polarized {K}\"ahler metrics on algebraic manifolds,"
J.\ Differential\ Geom.\ {\bf 32}, no.~1, 99-130 (1990)
doi:10.4310/jdg/1214445039.

\bibitem{Donaldson2}
S.~K. Donaldson, 
``Scalar curvature and projective embeddings {II},"
Q.\ J.\ Math.\ {\bf 56}, no.~3, 345-356 (2005)
doi:10.1093/qmath/hah044.

\bibitem{AtiyahBottGarding}
  M.~F.~Atiyah, R.~Bott, L.~Garding,
  ``Lacunas for hyperbolic differential operators with constant coefficients {II},"
  Acta.\ Math.\ {\bf 131}, 145-206 (1973)
  doi:10.1007/BF02392039

\bibitem{Candelas:1987kf} 
  P.~Candelas, A.~M.~Dale, C.~A.~Lutken and R.~Schimmrigk,
  ``Complete Intersection Calabi-Yau Manifolds,''
  Nucl.\ Phys.\ B {\bf 298}, 493 (1988).
  doi:10.1016/0550-3213(88)90352-5

\bibitem{Headrick:code}
Code associated to arXiv:0908.2635, which can be found at the following URL: \newline http://people.brandeis.edu/$\sim$headrick/Mathematica/index.html.
 

 \bibitem{Anderson:2017aux} 
  L.~B.~Anderson, X.~Gao, J.~Gray and S.~J.~Lee,
  ``Fibrations in CICY Threefolds,''
  JHEP {\bf 1710}, 077 (2017)
  doi:10.1007/JHEP10(2017)077
  [arXiv:1708.07907 [hep-th]].

  
  
   

  

  


  


  
    


\end{thebibliography}
\end{document}